\documentstyle[prabib,aps,preprint]{revtex}

\date{today}
\begin{document}
\noindent
\begin{center}{\Large \bf 
Important role of the spin-orbit interaction
in forming the $1/2^+$ orbital structure  in Be isotopes} \\

\vskip.25in
{\it N. Itagaki, S. Okabe$^*$, and K. Ikeda}
\address{itagaki@postman.riken.go.jp}

{\it RI Beam Science Laboratory, 
RIKEN (The Institute of Physical and Chemical Research), 
Wako, Saitama 351-0198, Japan}

{$^*$ \it Center for Information and Multimedia Studies,
Hokkaido University, \\
Sapporo 060-0810, Japan}
\end{center}

\begin{abstract}                
The structure of the second $0^+$ state
of $^{10}${Be} is investigated 
using a microscopic $\alpha$+$\alpha$+$n$+$n$ model
based on the molecular-orbit (MO) model.
The second $0^+$ state, which has dominantly 
the $(1/2^+)^2$ configuration,
is shown to have a particularly enlarged $\alpha$-$\alpha$ structure.
The kinetic energy of the two
valence neutrons occupying 
along the $\alpha$-$\alpha$ axis 
is reduced remarkably due to the strong $\alpha$ clustering
and, simultaneously, 
the spin-orbit interaction
unexpectedly plays important role
to make the energy of this state much lower.
The mixing of states with different spin structure 
is shown to be important in negative-parity states.
The experimentally observed small-level spacing between $1^-$ and $2^-$
($\sim$300 keV) is found to be an evidence of this spin-mixing effect. 
$^{12}${Be} is also investigated using $\alpha$+$\alpha$+$4n$ model,
in which four valence neutrons
are considered to occupy the $(3/2^-)^2(1/2^+)^2$ configuration.
The energy surface of $^{12}$Be
is shown to exhibit similar characteristics,
that the remarkable $\alpha$ clustering and 
the contribution of the spin-orbit interaction make the binding
of the state with $(3/2^-)^2(1/2^+)^2$ configuration
properly stronger in comparison with the closed $p$-shell
$(3/2^-)^2(1/2^-)^2$ configuration.
\end{abstract}
\begin{center}
PACS number(s): 21.10.-k, 21.60.Gx
\end{center}

\section{INTRODUCTION}
Numerous experiments using
unstable nuclear beams have succeeded to 
reveal exotic properties of $\beta$-unstable nuclei,
including a neutron halo structure\cite{Tani88,Nii,Tani96}.
Especially, the shift of the closed-shell structure is 
a characteristic behavior of systems with weakly bound neutrons.
Recently, the contributions of the $sd$ shell have been analyzed 
in $N=8$ nuclei based on the shell model.
A calculation has shown
that the slow $\beta$-decay of $^{12}$Be to $^{12}$B
can be explained by an admixture of the $sd$ shell in $^{12}$Be ($N=8$)
in which the closed $p$-shell component must be less than 
30$\%$\cite{Suzuki}. 
This shows that the concept of magic numbers is vague in $^{12}$Be.

On the other hand, an $\alpha$-$\alpha$ structure
is well established in the Be and B region.
Typically in $^9$Be,
a microscopic $\alpha$+$\alpha$+$n$ model has reproduced 
the properties of low-lying states
including the level inversion of $1/2^-$ in the $p$-shell
and $1/2^+$ in the $sd$ shell\cite{Okabe77,Okabe79}.
Also in $^{10}$Be, the microscopic 
$\alpha$-cluster models have been applied \cite{Okabe-S,Seya},
and we have quantitively shown the mechanism
for the lowering of the $sd$ orbits
related with the clustering of the core\cite{Ita}.

According to our previous results for $^{10}$Be,
the main properties of this nucleus have been well described by 
the $\alpha$+$\alpha$+n+n model, where the orbits for the
two valence neutrons are classified based on 
the molecular-orbit (MO) model.
The dominant configuration of the two valence 
neutrons for the second $0^+$ state is $(1/2^+)^2$,
and the level inversion between $1/2^-$ and $1/2^+$ in $^9$Be
also holds in $^{10}$Be.
Here, the two valence neutrons stay along the $\alpha$-$\alpha$ axis,
and reduce the kinetic energy by enhancing the
$\alpha$-$\alpha$ distance (up to around 4 fm).
Therefore, the $1/2^+$ orbit in this case is not a spherical
$s$-wave, but a polarized orbit with the $d$-wave component. 
This feature of $sd$ mixing in the $1/2^+$ orbit can be 
qualitatively interpreted in terms of deformed models 
([220] expression in the Nilsson diagram).

The main purpose of this paper is to show
the mechanism for the appearance of this $(1/2^+)^2$ configuration 
in low-lying energy in detail.
In our previous work, the calculated excitation energy of 
the second $0^+$ state has been higher by $\sim$5 MeV 
without the spin-orbit interaction
than the experimental one\cite{Ita}.
The spin-orbit interaction has been found to 
decrease the excitation energy by 3 MeV
by adding a single optimal wave function with $S=1$.
Hence, it is necessary to analyze the contribution of the
spin-orbit interaction more precisely to clarify
the lowering mechanism of the second $0^+$ state.
The other reduction 
mechanism of the kinetic energy
by more accurate description of the neutron-tail 
should also be taken into account.

Firstly, the spin-orbit interaction is able to
contribute to the second $0^+$ state, 
when spin-triplet states for the valence neutrons are included
among the basis states.
If the $1/2^+$ state consists of the pure s-orbit, naturally there is no 
contribution of the spin-orbit interaction.
In Be isotopes, the state contains the $d$-orbit component.
However, the spin-orbit interaction again vanishes,
when the two valence neutrons occupy along the $\alpha$-$\alpha$ axis,
since two neutrons with the same spatial distribution construct
only the spin-singlet state.
This has been the situation in traditional MO models\cite{Okabe-S,Seya}.
It will be shown that when one of the valence neutrons deviates
from the $\alpha$-$\alpha$ axis, the spin-triplet state can be 
constructed, and the spin-orbit interaction strongly acts 
between this state and the original $(1/2^+)^2$ configuration
with the spin-singlet.
The extension of the model space
also enables us to improve the description of the neutron-tail.
We estimate the reduction of the kinetic energy due to 
the improvement of the neutron-tail.
These improvements will be shown to largely decrease the excitation energy
of the second $0^+$ state, and that the calculated energy just corresponds
to the experimental value.

Using the present extended MO model,
negative-parity states in $^{10}$Be are also investigated. 
The mixing of different spin structure $(S=0, S=1)$ is found to 
also be important in the negative-parity states.
The $K$-mixing effect strongly acts on the $2^-$ state,
and the calculated small-level spacing between $1^-$ and $2^-$
($\sim$300 keV) agrees with the experimental value.

Also, in $^{12}$Be
these two important mechanism for lowering the $(1/2^+)^2$ configuration 
are examined. 
In $^{12}$Be, the configuration for four valence neutrons 
around the $\alpha$-$\alpha$ core is considered to correspond to 
the two pair configurations, $(3/2^-)^2$ and $(1/2^+)^2$. 
These two pair configurations
appear in the ground $0^+$ state 
and the second $0^+$ state of $^{10}$Be, respectively.
The energy surface for the ground state of $^{12}$Be will be shown,
in which the spin-orbit interaction is seen to make the binding
for the $(3/2^-)^2(1/2^+)^2$ configuration
properly stronger in comparison with the $(3/2^-)^2(1/2^-)^2$
configuration.
This is related to
the breaking of the $N=8$ (closed $p$-shell) neutron-magic number.
 
This paper is organized as follows. In Sec. II, we
summarize a description of the single-particle orbits around 
the two $\alpha$ clusters based on the MO model.
In Sec. III,
the contribution of the spin-orbit interaction 
to the $(1/2^+)^2$ configuration and the reduction
of the kinetic energy due to an improvement of the neutron-tail
are considered.
These effects are discussed in detail
for the second $0^+$ state of $^{10}$Be ({\bf A}), 
for the negative parity states of $^{10}$Be ({\bf B}),
and for $^{12}$Be ({\bf C}). 
The conclusion is given in Sec. IV.

\section{extended Molecular Orbital model} 
We introduce a microscopic $\alpha$+$\alpha$+$2n$ model for $^{10}$Be 
and $\alpha$+$\alpha$+$4n$ model for $^{12}$Be.
The neutron configurations are 
classified according to
the molecular-orbit (MO) picture\cite{Abe}.
The details of the framework are given in Ref.\cite{Ita}; here,
the essential part is presented.

The Hamiltonian is the same as in Ref.\cite{Ita}, 
and consists of a kinetic-energy term,
a central two-body interaction term, a two-body spin-orbit 
interaction term,
and a Coulomb interaction term: 
\begin{equation}
{\cal H}=\sum_i T_i - T_{cm} + \sum_{i < j} V_{ij}   
+ \sum_{i < j} V^{ls}_{ij} 
+ \sum_{i < j} {e^2 \over 4r_{ij}} (1-\tau_z^i)(1-\tau_z^j).
\end{equation}
The effective nucleon-nucleon interaction
is Volkov No.2\cite{VolkovInt} for the central part
and the G3RS spin-orbit term\cite{G3RS} 
for the spin-orbit part, as follows:
\begin{equation}
V_{ij} = \{V_1 e^{-a_1r_{ij}^2} - V_2 e^{-a_2r_{ij}^2}\} 
\{ W-MP^{\sigma}P^{\tau}+BP^{\sigma}-HP^{\tau} \},
\end{equation}
\begin{equation}
V^{ls}_{ij} = V^{ls}_0 \{ e^{-a_1r_{ij}^2}-e^{-a_2r_{ij}^2}\} 
{\vec L \cdot \vec S}P_{31},
\end{equation}
where $P_{31}$ is a projection
operator onto the triplet odd state.
The parameters are
$V_1 = -60.650$ MeV, $V_2 = 61.140$ MeV, $a_1 = 0.980$ fm$^{-2}$ 
and $a_2 = 0.309$ fm$^{-2}$ 
for the central interaction,
and $V^{ls}_0$ = 2000 MeV, 
$a_1 = 5.00$ fm$^{-2}$, and $a_2 = 2.778$ fm$^{-2}$
for the spin-orbit interaction.
We employ
the Majorana exchange parameter, $M = 0.6$ ($W = 0.4$),
the Bartlett exchange parameter, $B = 0.125$,
and the Heisenberg exchange parameter, $H = 0.125$, for the Volkov
interaction
(using $B$ and $H$, there is no neutron-neutron bound state).
All of these parameters are determined 
from the $\alpha+n$ and $\alpha+\alpha$ 
scattering phase shifts and the binding energy of
the deuteron\cite{Okabe79}.

The total wave function is fully antisymmetrized and
expressed by a superposition of terms
centered to different relative distances between 
the two $\alpha$ clusters ($d$)
with various configurations of the valence neutrons. 
The projection
to the eigen-states of angular momentum $(J)$ is performed numerically.
All nucleons are described by Gaussians 
with the oscillator parameter $(s)$ equal to 1.46 fm.
The $\alpha$ clusters located at $d/2$ and $-d/2$ 
on the $z$-axis consist of four nucleons:
\begin{equation}
\phi^{(\alpha)}=
G_{R_\alpha}^{p\uparrow}
G_{R_\alpha}^{p\downarrow}
G_{R_\alpha}^{n\uparrow}
G_{R_\alpha}^{n\downarrow}
\chi_{p\uparrow}
\chi_{p\downarrow}
\chi_{n\uparrow}
\chi_{n\downarrow}.
\end{equation}
$G$ represents Gaussians:
\begin{equation}
G_{R_\alpha} = \left( {2\nu \over \pi}\right)^{3 \over 4}
\exp[-\nu(\vec r-\vec R_\alpha)^2],
\ \ \ \ \ \ \nu=1/2s^2, 
\ \ \ \ \ \ \vec R_\alpha = \{ d\vec e_z/2, \ \ -d\vec e_z/2\}.
\end{equation}
Each valence neutron ($\phi_{ci}\chi_{ci}$) around 
the $\alpha$-$\alpha$ core
is expressed by a linear combination of local Gaussians:
\begin{equation}
\phi_{ci}\chi_{ci} =
\sum_{j} g_{j}
G_{R_n^j} 
\chi_{ci},
\end{equation}
\begin{equation}
G_{R_n^j}=\left( {2\nu \over \pi}\right)^{3 \over 4}
\exp[-\nu(\vec r-\vec R_n^j)^2],
\ \ \ \ \ \ \nu=1/2s^2. 
\end{equation}
The level structure is solved by superposing basis states
with different $\alpha$-$\alpha$ distances and configurations
after the angular momentum projection.

In the MO model, valence neutrons are expressed 
by a linear combination of orbits 
around two $\alpha$ clusters.
The orbit around each $\alpha$ cluster is called the atomic orbit (AO).
Here, we note that the valence neutrons and neutrons in
the $\alpha$ clusters are identical particles,
and antisymmetrization imposes the AO forbidden space.
The lowest AO has one node and parity minus, which is the $p$-orbit.
When a linear combination of two AOs ($p$-orbit) 
around the left-$\alpha$ cluster
and the right-$\alpha$ cluster is summed up by
the same sign,
the resultant MO also has negative parity and one node.
In this case, the MO is restricted to spread along 
an axis perpendicular 
to the $\alpha$-$\alpha$ $(z)$ axis, which is the so-called $\pi$-orbit. 
It cannot spread along the $z$-axis where 
the two $\alpha$ clusters are already located.
If two $p$-orbits are summed up by different signs, 
the resultant MO has two nodes and positive parity.
This MO can spread to all directions, 
and the optimal direction becomes the $z$-direction.
This is the so-called $\sigma$-orbit,
and the energy becomes lower as 
the distance between two $\alpha$ clusters
increases.

In Be isotopes, the low-lying orbits with 
$K^\pi=3/2^-$, $1/2^+$, and $1/2^-$ are 
important; these are classified based on the MO mode1.
The $1/2^+$ state is the $\sigma$-orbit, and the spin-orbit interaction 
splits the $\pi$-orbit to $3/2^-$ and $1/2^-$.
In the present framework,
each valence neutron is 
introduced to have a definite $K^\pi$
at the zero limit of centers of local Gaussians 
$(\{ R^j_n \})$ describing the spatial distribution of the orbit. 
The precise positions of $\{ R^j_n \}$
are determined variationally for each $\alpha$-$\alpha$ distance
before the angular-momentum projection.
Since the values of $\{ R^j_n \}$ are optimized to be finite,
the orbits are not exactly the eigen-state of $K^\pi$,
and are labeled as $\bar K^\pi$.

In $^{10}$Be, when two valence neutrons occupy 
these single particle orbits, three configurations
with the total $\bar K=\bar K_1+\bar K_2=0$ are generated. 
For the ground state,
valence neutrons occupy orbits with 
$\bar K^\pi = 3/2^-$ and $\bar K^\pi = -3/2^-$;
these $\pi$-orbits are expressed 
as linear combinations of $p_x$ and $p_y$:

\begin{equation}
\Phi({3/2}^-, {-{3/2}}^-) 
={\cal A}[\phi^{(\alpha)}_1\phi^{(\alpha)}_2
(\phi_{c1}\chi_{c1})(\phi_{c2}\chi_{c2})].
\end{equation}
\begin{equation}
\phi_{c1}\chi_{c1}= 
\{(p_x+ip_y)_{+a}+(p_x+ip_y)_{-a}\}|n\uparrow \rangle,
\end{equation}
\begin{equation}
\phi_{c2}\chi_{c2}= 
\{(p_x-ip_y)_{+a}+(p_x-ip_y)_{-a}\}|n\downarrow \rangle.
\end{equation}
Each MO is constructed from two AOs 
around a left and right $\alpha$ cluster.
We introduce
a variational parameter $(a)$ on the $\alpha$-$\alpha$ axis ($z$-axis);
and $(p_x\pm ip_y)_{+a}$ is 
$p_x\pm ip_y$, whose center is $+a$ on the $z$-axis;
and $(p_x\pm ip_y)_{-a}$ is
$p_x\pm ip_y$, whose center is $-a$ on the $z$-axis.
Here, the spatial part and the spin part of $\bar K$ in Eqs. (9) and (10)
are introduced to be parallel,
for which the spin-orbit interaction acts attractively.
The spin-up valence neutron 
has $\bar K^{\pi}_1=3/2^-$ ($rY_{11}|n\uparrow \rangle$), 
and the spin-down valence neutron 
has $\bar K^{\pi}_1=3/2^-$ ($rY_{1-1}|n\downarrow \rangle$).
There $(p_x)_{+a}$, $(p_x)_{-a}$, $(p_y)_{+a}$, and $(p_y)_{-a}$
are expressed as the combination of two Gaussians, whose centers are  
shifted by a variational parameter $(b)$ perpendicular to the $z$-axis.
For example,
\begin{equation}
(p_x)_{+a}=G_{a\vec e_z+b\vec e_x}-G_{a\vec e_z-b\vec e_x}, \ \ \
(p_y)_{+a}=G_{a\vec e_z+b\vec e_y}-G_{a\vec e_z-b\vec e_y}.
\end{equation}
These parameters $a$ and $b$, are optimized by using 
the Cooling Method in antisymmetrized
molecular dynamics (AMD) \cite{Enyo95,Enyo95b,Enyo95c,AMD,Ono92}
for each $\alpha$-$\alpha$ distance.

The $\sigma$-orbit pair, $(1/2^+)^2$, is prepared,
where each $\sigma$-orbit is expressed
by subtracting two AOs at $+a=d/2$ and $-a=-d/2$:
\begin{equation}
\Phi({1/2}^+, {-{1/2}}^+) 
={\cal A}[\phi^{(\alpha)}_1\phi^{(\alpha)}_2
(\phi_{c1}\chi_{c1})(\phi_{c2}\chi_{c2})],
\end{equation}
\begin{equation}
\phi_{c1}\chi_{c1}=
\{(\vec p)_{+a}-(\vec p)_{-a}\}|n\uparrow \rangle \ \ \ \ a=d/2,
\end{equation}
\begin{equation}
\phi_{c2}\chi_{c2}=
\{(\vec p)_{+a}-(\vec p)_{-a}\}|n\downarrow \rangle \ \ \ \ a=d/2,
\end{equation}
where,
\begin{equation}
(\vec p)_{\pm a}=(G_{+\vec b}-G_{-\vec b})_{\pm a}.
\end{equation}
Since the $\sigma$-orbit has two nodes, the direction of the orbit
is not limited due to the Pauli principle. 
It can thus take all directions.
Therefore, the direction and value of
parameter $\vec b$ for the center of the Gaussians are optimized.
As a result, it is optimized
to the direction of the $\alpha$-$\alpha$ axis.
This basis state shares the dominant component of the second $0^+$ 
state.

\section{RESULTS obtained using the extended MO model space}
\subsection{$J^\pi = 0^+$ states in $^{10}$Be}
Before giving the results, 
we summarize the calculated 
energies and the optimal $\alpha$-$\alpha$ distance 
for three configurations
($(3/2^-)^2$, $(1/2^-)^2$, and $(1/2^+)^2$)
discussed in Ref.\cite{Ita}.
In Fig. 1,
the $0^+$ energy curves of 
$\Phi({3/2}^-, {-3/2}^-)$, $\Phi({1/2}^+, {-1/2}^+)$,
and $\Phi({1/2}^-, {-1/2}^-)$ in $^{10}$Be
are presented as a function of the $\alpha$-$\alpha$ distance $(d)$.
These configurations are the dominant components of 
the first, second, and third $0^+$ states in $^{10}$Be,
respectively.

\noindent
\begin{center}
--------------\\
Fig. 1 \\
--------------\\
\end{center}
The energy of $\Phi({1/2}^+, {-1/2}^+$) (dominant component of  $0^+_2$) 
becomes lower as $d$ increases,
and the energy becomes minimum at an $\alpha$-$\alpha$
distance of $4 \sim 5$ fm.
However, it should be noted that the energy difference between
$\Phi({1/2}^+, {-1/2}^+)$ and $\Phi({3/2}^-, {-3/2}^-)$
in Fig. 1 is about 11 MeV. 
This is much larger than the experimental excitation energy
of the second $0^+$ state (6.26 MeV\cite{Ajz88}) by 5 MeV.


There are two reasons for this lack of 
binding energy by 5 MeV in the second $0^+$ state.
One is that 
the contribution of the spin-orbit interaction 
is not taken into account;
the other is that the tail of the 
valence neutron is not correctly described.
As for the spin-orbit interaction, in the traditional MO model,
the spin-orbit interaction has not contributed to
the $0^+$ state with this configuration. 
This is because, the two valence neutrons 
occupy the same spatial configurations with opposite spin direction,
and construct a spin-singlet state.
Since it is impossible for two valence neutrons 
with the same spin direction to occupy the same $\sigma$-orbit
along the $\alpha$-$\alpha$ axis,
the spin-triplet state is prepared
by making one of the valence neutrons  
deviate from the $\alpha$-$\alpha$ axis.
As for the tail effect of the valence neutrons, 
in this analysis,
the tail behavior is expressed by
the superposition of local Gaussians.

These two improvements are shown for 
the $\Phi({1/2}^+, {{-1/2}}^+)$ configuration
with the optimal $\alpha$-$\alpha$ distance of 4 fm.
The optimized parameter $\vec b$ 
is $2.02 \vec e_z$ (fm)\cite{Ita}.
As listed in Table I,
the spin-singlet $\Phi({1/2}^+, {{-1/2}}^+)$ configuration 
has a $0^+$ energy of $-46.3$ MeV. 
As shown in Eqs. (12)-(15),
each valence neutron is described as a linear combination
of four local Gaussians. When we optimize their linear combination,
the energy slightly decreases to $-47.7$ MeV (noted as
$S=0$ $(\sigma)^2$ in Table I).

\noindent
\begin{center}
--------------\\
Table I \\
--------------\\
\end{center}
Next, 
the contribution of the spin-orbit interaction 
is taken into account to this spin-singlet state
by preparing
the spin-triplet state ($S_z = 1$). 
One of the valence neutrons ($\phi_{c1}\chi_{c1}$)
occupies the $\sigma$-orbit, and
another one ($\phi_{c2}\chi_{c2}$) with the same spin-direction 
deviates from the $\alpha$-$\alpha$ axis.
The later one is
described by a local Gaussian centered at $\vec R$;
$\phi_{c2}\chi_{c2}=G_{\vec R}|n\uparrow \rangle$.
The calculated energy surface for the second $0^+$ state
as a function of the parameter $\vec R$ 
is given in Fig. 2.
The orthogonality of the ground $0^+$ state
to the ground $0^+$ state is ensured in this calculation.
The point on the $x$-$z$ plain in Fig. 2 shows the position of 
the parameter $\vec R$, and the contour map 
shows the total energy obtained by taking into account this
coupling with the spin-triplet states.

\noindent
\begin{center}
-------------\\
Fig. 2 \\
-------------\\
\end{center}
The lowest energy of $-50.9$ MeV is seen 
at $\vec R = (x,z) = (1.0,2.0)$ (fm) on the energy surface.
The contribution of the spin-orbit interaction
decreases the energy by more than 3 MeV in comparison with 
the case of the spin-singlet state.
Although the energy of such a spin-triplet state, itself, is 
much higher than that of $\Phi({1/2}^+, {{-1/2}}^+)$ by about 20 MeV,
the large coupling between the spin-singlet and spin-triplet states
enables the spin-orbit interaction 
to attractively act.
As a result, more than half of the under-binding
by 5 MeV 
for the second $0^+$ state 
is explained by this spin-orbit contribution. 

To see this effect more correctly,
a lot of spin-triplet states with different $\vec R$ values
are adopted.
The employed $\vec R$ values are 0.0, 1.0, 2.0 (fm)
for the $x$-direction
(perpendicular to the $\alpha$-$\alpha$ direction)
and 0.0, 1.0, 2.0, 3.0, 4.0 (fm) for the $z$-direction 
(the $\alpha$-$\alpha$ direction).
In Table I, the column noted as spin-orbit gives the energy
of the second $0^+$ state when 
these 15 spin-triplet states with different $\vec R$ values are superposed.
The calculated energy is $-51.2$ MeV; this is lower than the energy minimum
point in Fig. 2 by only 300 keV. This is because,
the spin-triplet basis states around the core have a large overlap 
with each other, and those 
apart from the core do not contribute to this coupling.
This result shows that 
the contribution of the spin-orbit interaction for the 
$(1/2^+)^2$ state is  
approximately taken into account when we employ
only one Slater determinant on the energy surface.

In this calculation, many spin-triplet states are employed;
however, it should be noted that
the spin-singlet state is only
represented by the $\Phi({1/2}^+, {{-1/2}}^+)$ configuration.
Thus, to improve the tail of the valence neutron 
in the second $0^+$ state of $^{10}$Be,
it is necessary to superpose further the $S_z=0$ basis states 
with many different $\vec R$ values,
which contain the spin-singlet component. 
The distance of $\alpha$-$\alpha$ is 4 fm,
where one of the valence neutrons ($\phi_{c1}\chi_{c1}$) occupies the 
$\sigma$-orbit, and
the other valence neutrons ($\phi_{c2}\chi_{c2}$) with the opposite
spin-direction is described by local Gaussian centered at $\vec R$.
The later one has the form of
$\phi_{c2}\chi_{c2}=G_{\vec R}|n\downarrow \rangle$.
We superpose 15 basis states with $S_z=0$,
and the employed $\vec R$ values are 0.0, 1.0, 2.0 (fm)
for the $x$-direction
(perpendicular to the $\alpha$-$\alpha$ direction)
and 0.0, 1.0, 2.0, 3.0, 4.0 (fm) for the $z$-direction
(the $\alpha$-$\alpha$ direction).
As listed in Table I, 
the energy becomes $-52.7$ MeV.
These basis states increase the binding of the 
second $0^+$ state by 1.5 MeV compared with $-51.2$ MeV for the third row.
Here, it is noted that these $S_z = 0$ basis states
contain both the spin-single and spin-triplet components.
However, the increase of the binding energy 
by adding these $S_z = 0$ basis states
is due to the spin-singlet basis states,
since it is considered that
the functional space of the spin-triplet states are already exhausted 
by the $S_z = 1$ states.
These results show that
by taking into account both the spin-orbit interaction (spin-triplet
basis states) and the tail effect (spin-singlet basis states),
the under-binding for the second $0^+$ state by 5 MeV
is fully explained.
The calculated excitation energy of the 
second $0^+$ state agrees with the experimental one.

To confirm the component of the spin-triplet configurations
introduced,
the expectation value of the spin-square $(\hat S^2$) 
in the second $0^+$ state of $^{10}$Be was calculated. 
The value is 0.26.
This means that the spin-triplet configuration mixes in this state
by about 13$\%$.

In this analysis, the $\alpha$-$\alpha$ distance
is restricted to 4 fm,
but
we now examine the contribution of the spin-orbit interaction with respect to 
the $\alpha$-$\alpha$ distance.

\noindent
\begin{center}
--------------\\
Table II \\
--------------\\
\end{center}
Table II gives the contribution of spin-orbit interaction
for the second $0^+$ state
with the $\alpha$-$\alpha$ distance of 3 fm, 4 fm, and 5 fm.
The column noted as $S=0$ $(\sigma)^2$ gives the energy
for the $\Phi({1/2}^+, {{-1/2}}^+)$ configuration, 
where the linear combination
of Gaussians is optimized.
The column noted as $+S_z = 1$ gives the energy 
when we include the spin-orbit interaction
by employing the spin-triplet basis states.
The results show that the smaller is the $\alpha$-$\alpha$ distance, 
the larger is
the contribution of the coupling with the spin-triplet state.
When the $\alpha$-$\alpha$ distance is 5 fm, 
the coupling increases the binding energy by about 3 MeV; however, 
when the $\alpha$-$\alpha$ distance is 3 fm, the coupling increases
the binding energy by about 4.5 MeV. 
Therefore, the coupling with the spin-triplet 
becomes stronger as the $\alpha$-$\alpha$ distance becomes small.
The column noted as $+S_z = 0,1$ gives the energy 
when we further include the spin-singlet basis state to 
express the tail of the valence neutron.
This effect does not have a strong dependence 
on the $\alpha$-$\alpha$ distance, and almost constantly
decreases the binding energy by 1.5 MeV. 

Finally, the calculation where
all of these basis states with different
$\alpha$-$\alpha$ distance are superposed 
is performed based on the
generator coordinate method (GCM).
When we superpose all of the basis states in Table II,
the calculated energy of the second $0^+$ state becomes
$-54.3$ MeV. Here, we superpose the basis states
with $\alpha$-$\alpha$ distances of
3, 4, 5, and 6 fm,
and the spin-triplet states are represented
by one Slater determinant for each $\alpha$-$\alpha$ distance,
which correspond to the minimum point on the energy surface (Fig. 2).
This calculated energy of $-54.3$ MeV is lower than a result 
calculated with only 
the $\Phi({1/2}^+, {{-1/2}}^+)$ configuration (employed $\alpha$-$\alpha$
distances are the same) by 4.6 MeV.


\subsection{Negative-parity states in $^{10}$Be}
We examine this spin-mixing effect for
the negative-parity states in $^{10}$Be.
Here, one of the valence neutrons is introduced to occupy
a $\pi$-orbit with negative parity;
another one occupies a $\sigma$-orbit with positive parity.
In this way, two configurations with 
$S_z = 0$ and $S_z = 1$ can be constructed.
The $S_z = 0$ basis state ($\Phi({3/2}^-, {{-1/2}}^+)$)  
has $K^\pi = 1^-$. 
\begin{equation}
\Phi({3/2}^-, {{-1/2}}^+)
={\cal A}[\phi^{(\alpha)}_1\phi^{(\alpha)}_2
(\phi_{c1}\chi_{c1})(\phi_{c2}\chi_{c2})].
\end{equation}
\begin{equation}
\phi_{c1}\chi_{c1}= 
\{(p_x+ip_y)_{+a}+(p_x+ip_y)_{-a}\}|n\uparrow \rangle,
\end{equation}
\begin{equation}
\phi_{c2}\chi_{c2}=
\{(\vec p)_{+a}-(\vec p)_{-a}\}|n\downarrow \rangle.
\end{equation}
This basis state gives $-47.7$ MeV for $1^-$ and $-46.7$ MeV for $2^-$
at the optimal $\alpha$-$\alpha$ distance of 3 fm.
This energy splitting between $1^-$ and $2^-$ by 1 MeV
is much larger than the experimental value of 300 keV
(experimentally, the excitation energies of the $1^-$ state 
and the $2^-$ state are 5.96 MeV and 6.26 MeV, respectively).
At first, we only examine
the tail effect of the valence neutrons with $S_z = 0$.
We introduce a parameter $\vec R$ for 
$\phi_{c2}\chi_{c2}$,
which originally occupies the $\sigma$-orbit:
\begin{equation}
\phi_{c2}\chi_{c2}=G_{\vec R}|n\downarrow \rangle. 
\end{equation}
However, even if we superpose states with many different positions
($\vec R$), the energy splitting between these states is still about 1 MeV
($1^-$ state is $-49.9$ MeV and the $2^-$ state is $-48.9$ MeV).
The employed $\vec R$ values are 0.0, 1.0, 2.0 (fm)
for the $x$-direction (perpendicular to the $\alpha$-$\alpha$ direction)
and 0.0, 1.0, 2.0, 3.0, 4.0 (fm) for the $z$-direction ($\alpha$-$\alpha$
direction).

Therefore, to improve this discrepancy, we next
include the basis states with $S_z = 1$; $\Phi(3/2^-, 1/2^+)$.
The $S_z = 1$ basis state
generates $K^\pi = 2^-$ band.
As a result of 
the calculation, the energy of the $2^-$ state
with $K^\pi=2^-$ (spin triplet) is found to be
accidentally close to the energy of the
$2^-$ state with $K^\pi = 1^-$
projected from the $S_z = 0$ basis states.
Thus, the coupling effect between the $S_z = 0$ basis states
and the $S_z = 1$ basis states is larger for the $2^-$ state 
than for the $1^-$ state, and the energy splitting becomes 
comparable to the experimental one.
To see this effect, we superpose many $S_z = 1 $ basis states 
characterized by the Gaussian center $\vec R$;
$\phi_{c2}\chi_{c2}=G_{\vec R}|n\downarrow \rangle$.
When we include many states with different $\vec R$ values
in our bases states,
the energies become $-51.6$ MeV ($1^-$) and 
$-51.2$ MeV ($2^-$).
The employed $\vec R$ values are 0.1, 1.0,  2.0 (fm)
for the $x$-direction
and 0.0, 1.0, 2.0, 3.0, 4.0 (fm) for the $z$-direction.
The experimental small level splitting 
between $1^-$ and $2^-$ 
is found to be evidence of the spin vibration
in the $2^-$ state.

\subsection{The structure of $^{12}$Be}
The large contribution of this spin-orbit interaction
will be discussed concerning $^{12}$Be.
In the previous subsection, 
we have discussed for $^{10}$Be that this effect is more important
as the $\alpha$-$\alpha$ becomes smaller, and in $^{12}$Be,
we show that the distance is smaller than that of the second 
$0^+$ state of $^{10}$Be.
In $^{12}$Be, four valence neutrons rotate around two $\alpha$ clusters
and, mainly, two configurations are important for the $0^+$ 
ground state\cite{Icc99}.
One is $(3/2^-)^2(1/2^-)^2$ for the four valence neutrons,  
which corresponds to the closed $p$-shell configuration of the neutrons
at the zero limit of the $\alpha$-$\alpha$ distance.
The other configuration is $(3/2^-)^2(1/2^+)^2$, 
where two of the four valence neutrons
occupy the $\sigma$-orbit. 
It is necessary to compare the energy of these two configurations as
a function of the $\alpha$-$\alpha$ distance.
The configurations
of $(3/2^-)^2(1/2^-)^2$ and
$(3/2^-)^2(1/2^+)^2$ are constructed 
as follows:
\begin{equation}
\Phi(3/2^-, -3/2^-, -1/2^-, 1/2^-) 
={\cal A}[\phi^{(\alpha)}_1\phi^{(\alpha)}_2
(\phi_{c1}\chi_{c1})(\phi_{c2}\chi_{c2})
(\phi_{c3}\chi_{c3})(\phi_{c4}\chi_{c4})]
\end{equation}
\begin{equation}
\phi_{c1}\chi_{c1}= 
\{(p_x+ip_y)_{+a}+(p_x+ip_y)_{-a}\}|n\uparrow \rangle,
\end{equation}
\begin{equation}
\phi_{c2}\chi_{c2}= 
\{(p_x-ip_y)_{+a}+(p_x-ip_y)_{-a}\}|n\downarrow \rangle,
\end{equation}
\begin{equation}
\phi_{c3}\chi_{c3}=
\{(p_x-ip_y)_{+a}+(p_x-ip_y)_{-a}\}|n\uparrow \rangle,
\end{equation}
\begin{equation}
\phi_{c4}\chi_{c4}=
\{(p_x+ip_y)_{+a}+(p_x+ip_y)_{-a}\}|n\downarrow \rangle,
\end{equation}
\begin{equation}
\Phi(3/2^-, -3/2^-, 1/2^+, -1/2^+) 
={\cal A}[\phi^{(\alpha)}_1\phi^{(\alpha)}_2
(\phi_{c1}\chi_{c1})(\phi_{c2}\chi_{c2})
(\phi_{c3}\chi_{c3})(\phi_{c4}\chi_{c4})]
\end{equation}
\begin{equation}
\phi_{c1}\chi_{c1}= 
\{(p_x+ip_y)_{+a}+(p_x+ip_y)_{-a}\}|n\uparrow \rangle,
\end{equation}
\begin{equation}
\phi_{c2}\chi_{c2}= 
\{(p_x-ip_y)_{+a}+(p_x-ip_y)_{-a}\}|n\downarrow \rangle,
\end{equation}
\begin{equation}
\phi_{c3}\chi_{c3}=
\{(\vec p)_{+a}-(\vec p)_{-a}\}|n\uparrow \rangle \ \ \ \ a=d/2,
\end{equation}
\begin{equation}
\phi_{c4}\chi_{c4}=
\{(\vec p)_{+a}-(\vec p)_{-a}\}|n\downarrow \rangle \ \ \ \ a=d/2.
\end{equation}

The dotted line in Fig. 3 shows $0^+$ energies of 
$(3/2^-)^2(1/2^+)^2$ and the dashed line shows $(3/2^-)^2(1/2^-)^2$ 
with respect to the $\alpha$-$\alpha$ distance.

\noindent
\begin{center}
-------------\\
Fig. 3 \\
-------------\\
\end{center}
When the $\alpha$-$\alpha$ distance is small, for example 2 fm, 
the dominant configuration of the four valence neutrons
is $(3/2^-)^2(1/2^-)^2$ for the ground state,
which corresponds to the closed $p$-shell configuration 
at the limit of the $\alpha$-$\alpha$ distance, zero.
On the other hand,
the $(3/2^-)^2(1/2^+)^2$ configuration for the four valence neutrons 
becomes lower as the $\alpha$-$\alpha$ distance is increased.
However, it is still higher than the
$(3/2^-)^2(1/2^-)^2$ configuration by a few MeV.
Here, we show the importance of the coupling 
between the $(3/2^-)^2(1/2^+)^2$ configuration  
and the spin-triplet state for the last two neutrons ($(1/2^+)^2$).
The spin-triplet state for $(1/2^+)^2$ is introduced as follows:
\begin{equation}
\Phi(3/2^-, -3/2^-, 1/2^+, G_{\vec R}|\uparrow \rangle)
={\cal A}[\phi^{(\alpha)}_1\phi^{(\alpha)}_2
(\phi_{c1}\chi_{c1})(\phi_{c2}\chi_{c2})
(\phi_{c3}\chi_{c3})(\phi_{c4}\chi_{c4})]
\end{equation}
\begin{equation}
\phi_{c1}\chi_{c1}= 
\{(p_x+ip_y)_{+a}+(p_x+ip_y)_{-a}\}|n\uparrow \rangle,
\end{equation}
\begin{equation}
\phi_{c2}\chi_{c2}= 
\{(p_x-ip_y)_{+a}+(p_x-ip_y)_{-a}\}|n\downarrow \rangle,
\end{equation}
\begin{equation}
\phi_{c3}\chi_{c3}=
\{(\vec p)_{+a}-(\vec p)_{-a}\}|n\uparrow \rangle \ \ \ \ a=d/2,
\end{equation}
\begin{equation}
\phi_{c4}\chi_{c4}=G_{\vec R}|n\uparrow \rangle.
\end{equation}
The coupling between the $(3/2^-)^2(1/2^+)^2$ configuration  
and the spin-triplet states is shown in Fig. 4 as a function 
of the parameter $\vec R$.

\noindent
\begin{center}
-------------\\
Fig. 4 \\
-------------\\
\end{center}
The point on the $x$-$z$ plain in Fig. 4 shows 
the Gaussian-center $\vec R$ in the spin-triplet state.
The contour map 
shows the energy calculated by taking into account this coupling
with the spin-triplet state. 
This $\alpha$-$\alpha$ distance is 3 fm, and this 
optimal distance is smaller than 
that for the second $0^+$ state of $^{10}$Be by about 1 fm, 
since two of the four valence neutrons occupy 
the $(3/2^-)^2$ configuration and increase the binding 
energy of the system. 
Because of this effect, the coupling effect
of the spin-triplet states is more important 
than in the case of the second $0^+$ state of $^{10}$Be.
The minimal point of the surface
shows that the energy gain due to the couplings is about 4 MeV in $^{12}$Be.

The solid line in Fig. 3 shows
the $0^+$ energy calculated by taking into account
the coupling effect
between the $(3/2^-)^2(1/2^+)^2$ configuration
and the $S_z=0,1$ basis states for the last two valence neutrons.
Due to the spin-orbit coupling, the energy is almost the same as that of
$(3/2^-)^2(1/2^-)^2$
corresponding to the closed $p$-shell configuration.
Furthermore, the energy of
$(3/2^-)^2(1/2^+)^2$ is suggested to become
even lower than $(3/2^-)^2(1/2^-)^2$
when the pairing effect between $(3/2^-)^2$ and $(1/2^-)^2$
is taken into account.
These effects
plays crucial roles in accounting for 
breaking of the $N = 8$ magic number.

\section{CONCLUSION}
The second $0^+$ state of $^{10}$Be
has been shown to be
characterized by the $\sigma$-orbit of the two valence neutrons
in terms of the molecular orbit (MO) model.
The two valence neutrons 
stay along the $\alpha$-$\alpha$ axis (the $1/2^+$ orbit)
and reduce the kinetic energy by enlarging the
$\alpha$-$\alpha$ distance.
This simple description for the second $0^+$
state gives a higher excitation energy by 5 MeV
compared to the experimental one.
To improve the description, 
spin-triplet basis states
which have not been included
in the traditional MO models are prepared by allowing
a deviation of 
the neutron orbit from one just along the $\alpha$-$\alpha$ axis.
Thus, the spin-orbit interaction can be taken into account,
and the calculated second $0^+$ state becomes lower by 3.5 MeV.
The precise description of the neutron-tail
also decreases the excitation energy by 1.5 MeV.
The discrepancy between the experimental excitation energy
and the calculated one is compensated by these two effects.

The spin-mixing effects were studied for
the negative parity state of $^{10}$Be.
If we restrict ourselves to the $S_z = 0$ basis state,
this energy splitting between the $1^-$ state and $2^-$ state
is about 1 MeV, which
is much larger than the experimental value of 303 keV.
As a result of adding basis states with $S_z = 1$, where
the spin directions of the two valence neutrons are the same,
the $K$-mixing occurs especially for the $2^-$ state.
The contribution of the $S_z = 1$ state is larger for the $2^-$ 
state. The resultant energy splitting becomes comparable to the
experimental one.
Therefore, the experimental small level splitting 
between $1^-$ and $2^-$ 
is considered to be a result of the spin vibration
induced by the spin-orbit interaction in the $2^-$ state.

The coupling 
with the spin-triplet basis states is also important in the case 
of $^{12}$Be.
Without the spin-triplet basis state,
the energy of the configuration $(3/2^-)^2 (1/2^+)^2$ is 
much higher than that of the closed $p$-shall configuration
($(3/2^-)^2 (1/2^-)^2$) by 4 MeV.
However, the energy of $(3/2^-)^2 (1/2^+)^2$
is drastically decreased by
coupling with the spin-triplet states.
This is because the effect
becomes stronger as the $\alpha$-$\alpha$ distance becomes shorter,
and $^{12}$Be has an optimal $\alpha$-$\alpha$
distance around 3 fm, which is smaller 
than the second $0^+$ state of $^{10}$Be by 1 fm.
The study shows that
an energy of $(3/2^-)^2 (1/2^+)^2$ is almost the same as
$(3/2^-)^2 (1/2^-)^2$, or even lower.
This effect is suggested to play a crucial
role in accounting for the dissipation
of the $N=8$ magic number in $^{12}$Be.
It is an interesting subject to analyze the 
binding mechanism and properties of
the ground state by taking into account the pairing mixing
among states with configurations of
$(3/2^-)^2 (1/2^-)^2$, $(3/2^-)^2 (1/2^+)^2$,
and  $(1/2^-)^2 (1/2^+)^2$.
A detail analysis is going to be performed not only
for this state, but also for excited states where
new states with cluster structure have been recently observed.

\acknowledgments

The authors would like to thank
Prof. I. Tanihata for various effective suggestions.
They also thank other members of RI beam science laboratory
in RIKEN for discussions and encouragements.
One of the authors (N.I) thanks 
fruitful discussions with Prof. R. Lovas,
Prof. H. Horiuchi, Prof. T. Otsuka, Prof. Y. Abe, Prof. K. Kat\=o, 
Prof. K. Yabana, Dr. A. Ohnishi, and Dr. Y. Kanada-En'yo.

\begin{figure}
\caption{
The $0^+$ energy curve for 
$\Phi({3/2}^-, {-3/2}^-$), $\Phi({1/2}^-, {-1/2}^-$),
and $\Phi({1/2}^+, {-1/2}^+$)
as a function of the $\alpha$-$\alpha$ distance $(d)$.
}
\end{figure}
\begin{figure}
\caption{
The energy surface of the second $0^+$ state 
of $^{10}$Be.
The total energy including
the coupling effect
between the $(1/2^+)^2$ configuration
and the spin-triplet configuration is shown. 
The $\alpha$-$\alpha$ distance is optimal 4 fm,
and the spheres show $\alpha$ clusters.
The axes (fm) show the position of
the Gaussian-center of one valence neutron in the 
spin-triplet state
which deviates from the $\sigma$-orbit.
}
\end{figure}
\begin{figure}
\caption{
The energy surface of 
the ground state of $^{12}$Be.
The total energy including
the coupling effect
between the $(3/2^-)^2(1/2^+)^2$ configuration
and the spin-triplet configuration for the last
two valence neutrons is shown. 
The $\alpha$-$\alpha$ distance is optimal 3 fm,
and the spheres show $\alpha$ clusters.
The axes (fm) show the position of
the Gaussian-center of one valence neutron in the 
spin-triplet state
which deviates from the $\sigma$-orbit
}
\end{figure}

\begin{figure}
\caption{
The $0^+$ energies for $(3/2^-)^2(1/2^+)^2$ configuration
(dotted line)
and $(3/2^-)^2(1/2^-)^2$ configuration (dashed line) in $^{12}$Be.
In the solid line, the couplings among
the $(3/2^-)^2(1/2^+)^2$ configuration,
an optimal $S_z=1$ state,
and $S_z=0$ states are included.
The resultant state almost degenerates with the
 $(3/2^-)^2(1/2^-)^2$ configuration.
}
\end{figure}

\begin{table}
\caption{The energy convergence of the second $0^+$ of $^{10}$Be.
The $\alpha$-$\alpha$ distance is fixed to 4 fm.
The row noted as $\Phi({1/2}^+, {{-1/2}}^+)$
shows the energy of the $\Phi({1/2}^+, {{-1/2}}^+)$ configuration, 
and the row $S=0$ $(\sigma)^2$ shows the energy
when a linear combination of local Gaussians is optimized.
The row spin-orbit shows the energy 
including the coupling with the spin-triplet state.
The row tail shows the calculated energy when
the tail of the valence neutron is further expressed by 
the $S_z=0$ basis states.
The values in parentheses give the energy differecne 
with the previous row. }

\begin{center}
\begin{tabular}{|c|c|c|}
 & number of Gaussians for one valence neutron & $0^+$ energy \\ 
\hline
$\Phi({1/2}^+, {{-1/2}}^+)$   &  4  & -46.3 \\ 
$S=0$ $(\sigma)^2$            &  4  & -47.7 (-1.4) \\ 
spin-orbit                    & 24  & -51.2 (-3.5) \\
tail                          & 44  & -52.9 (-1.7) \\
\end{tabular}
\end{center}
\end{table}

\begin{table}
\caption{The second $0^+$ energy of $^{10}$Be 
for each $\alpha$-$\alpha$ distance.
The row $S=0$ $(\sigma)^2$ shows the energy
of the $\Phi({1/2}^+, {{-1/2}}^+)$ configuration, 
where a linear combination
of local Gaussians is optimized.
The row $+S_z = 1$ represents the energy including the coupling 
between $S=0$ $(\sigma)^2$ 
and the spin-triplet state.
The row $+S_z = 0,1$ represents the energy including the coupling
between $S=0$ $(\sigma)^2$ 
and both the spin-singlet and the spin-triplet states.
The values in parentheses give the energy differecne 
with the $S=0$ state. }
\begin{center}
\begin{tabular}{|c|c|c|c|}
$\alpha$-$\alpha$ distance (fm) & 
 $S=0$ $(\sigma)^2$ (MeV) & $+S_z = 1$ (MeV) & $+S_z = 0,1$ (MeV) \\
\hline
3   & -44.5 & -48.9 (-4.4) & -50.6 (-6.1)\\
4   & -47.7 & -51.2 (-3.5) & -52.9 (-5.0)\\
5   & -46.9 & -50.0 (-3.1) & -51.5 (-4.6)\\
\end{tabular}
\end{center}
\end{table}

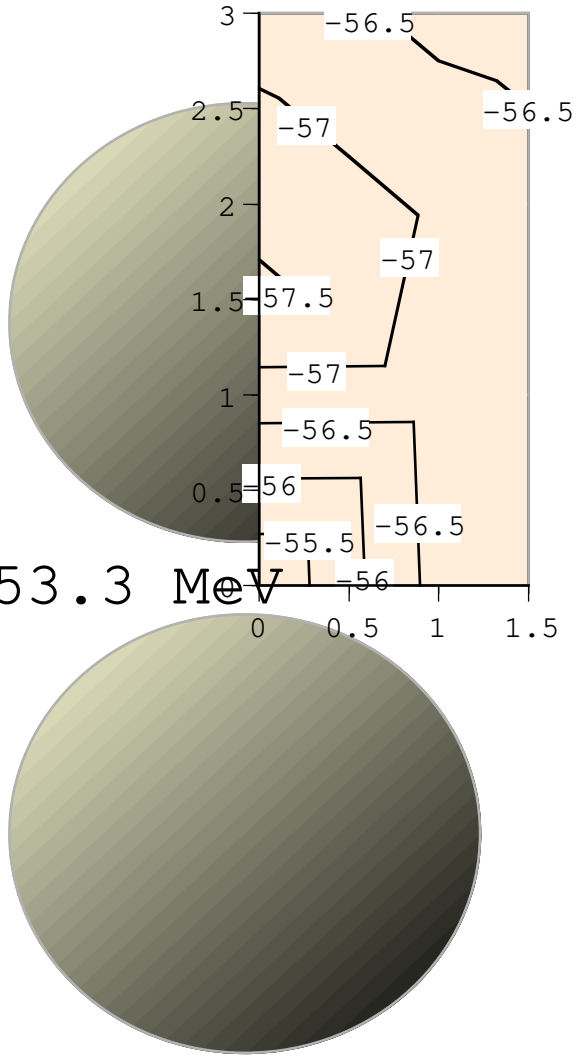
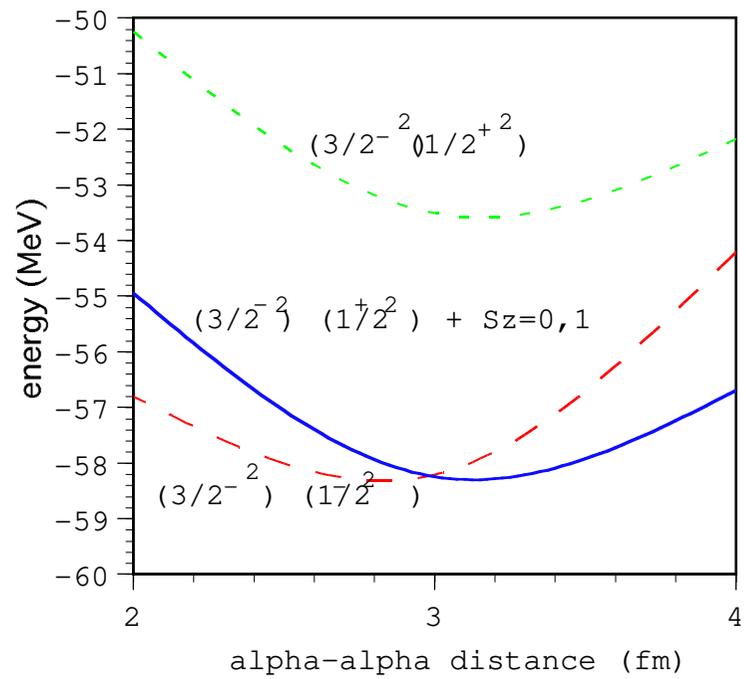
\end{document}